\documentclass[journal]{IEEEtran}
\usepackage{ifpdf}
\usepackage{cite}
\ifCLASSINFOpdf
   \usepackage[pdftex]{graphicx}
\else
\fi
\usepackage{amsmath}
\usepackage{algorithmic}
\usepackage{array}
\usepackage{amsmath,amssymb,bm}
\usepackage{tikz}
\usetikzlibrary{arrows.meta,positioning,fit,calc}

\begin{document}
\title{Near-field Channel Estimation for XL-RIS-aided  mmWave MIMO Systems}

\author{Erkang Dong, Taihao Zhang, Cunhua Pan, Hong Ren and Jiangzhou Wang
\thanks{The authors are with the National Mobile Communications Research Laboratory, Southeast University, Nanjing 210096, China (e-mail: 18451995211@163.com; taihao@seu.edu.cn; cpan@seu.edu.cn; hren@seu.edu.cn; j.z.wang@seu.edu.cn).}}






\maketitle 

\begin{abstract}
Extremely large-scale reconfigurable intelligent surfaces (XL-RISs) have emerged as a promising technology for millimeter-wave (mmWave) communications. However, the exceedingly large aperture of XL-RISs renders the RIS-user links likely to operate in the near-field region, where the conventional planar-wave assumption and angular-domain sparse representation become invalid, thus making channel estimation significantly more challenging. In this paper, we investigate cascaded channel estimation for an XL-RIS-aided multi-user multiple-input multiple-output (MU-MIMO) system, in which the BS-RIS channel is modeled in the far field, while the RIS-user channels exhibit near-field spherical-wave characteristics. To tackle the resulting hybrid-field estimation problem, we propose a low-overhead two-stage channel estimation scheme by jointly exploiting the common BS-RIS link shared by all users and the polar-domain sparsity of the RIS-user channels. Specifically, the multi-antenna users are firstly decomposed into multiple virtual single-antenna users, based on which the common BS-RIS parameters are extracted from a typical virtual user and the RIS-user channels are initialized via compensated polar-domain sparse recovery. Then, an alternating least-squares refinement procedure is developed to jointly improve the common BS-RIS operator and the user-specific RIS-side channels. Simulation results show that the proposed scheme achieves competitive channel estimation performance with substantially reduced pilot overhead compared with the existing near-field benchmarks.
\end{abstract}


% \fi

% ====================================================================
% ====================================================================
% ====================================================================

% === I. INTRODUCTION =============================================================
% =================================================================================
\section{Introduction}

\IEEEPARstart{T}{he} recent years have witnessed the evolving trends in wireless communication that reconfigurable intelligent surface(RIS) has been extensively researched. In particular, by properly tuning the passive reflecting elements, RISs are able to enhance the received signal strength and improve the coverage performance of wireless systems. Motivated by these appealing advantages, the power scaling law and phase-shift optimization of RIS-aided massive MIMO systems were investigated in \cite{IEEEhowto:zhi}. Meanwhile, the vision of smart radio environments enabled by RISs has been comprehensively discussed in \cite{IEEEhowto:DiRenzo}, and the fundamental principles, typical applications, and future research directions of RIS-assisted communications were further summarized in \cite{IEEEhowto:pan}. These studies have laid a solid foundation for the development of RIS-aided wireless systems.

With the rapid evolution of extremely large-scale MIMO (XL-MIMO) and XL-RIS systems, the near-field propagation effect has attracted increasing attention. The work in \cite{IEEEhowto:Dai} revealed that channel estimation for extremely large-scale MIMO should fundamentally distinguish between far-field and near-field regimes, which highlighted the necessity of polar-domain modeling for near-field channels. Moreover, the performance characteristics of extremely large-scale MIMO communications systems were further analyzed in \cite{IEEEhowto:lianXL-MIMO}, which also demonstrated the unique propagation behaviors introduced by extra-large array apertures. In such systems, the array aperture becomes sufficiently large such that the Rayleigh distance is significantly enlarged, and practical users may lie within the near-field region of the array. Therefore, the near-field effect brought about by XL-RIS needs to be considered. In the scenario of RIS-assisted communication in the far field, the common BS-RIS channel shared by all users and cascaded channel structure were exploited to reduce the pilot overhead in \cite{IEEEhowto:zhou}, and the corresponding framework was further extended to RIS-aided multiuser mmWave systems with uniform planar arrays in \cite{IEEEhowto:peng} and communication scenario assisted by RIS with direct path in \cite{IEEEhowto:zhang}. However, there are no related works on channel estimation for the near-field RIS assisted communication scenario. The authors in \cite{IEEEhowto:songjieyang} studied channel estimation for near-field XL-RIS-aided mmWave communication, where the importance of exploiting near-field structured sparsity was clearly verified, but the features of RIS were not fully utilized. In addition, more flexible RIS deployment scenarios, such as aerial RIS-assisted systems, may introduce further channel variations and impairments, as shown in \cite{IEEEhowto:lianARIS}, which makes reliable channel acquisition even more important.

This paper focuses on channel estimation for near-field RIS-aided multiuser MIMO systems. 
\textit{First,} we establish a cascaded channel estimation framework in which the BS-RIS link is characterized by a common far-field structured model, whereas the RIS-user links are modeled by near-field spherical-wave propagation with polar-domain sparsity. 
\textit{Second,} by jointly exploiting the common BS-RIS structure and the polar-domain sparse representation of the RIS-user channels, we develop a low-overhead channel estimation strategy for multiuser systems. 
\textit{Third,} the resulting design is particularly appealing for near-field XL-RIS-assisted mmWave communications, where conventional far-field estimation methods may encounter severe performance degradation caused by model mismatch.

\section{Channel Models}

\subsection{Channel Model}

We consider an XL-RIS-aided narrow-band TDD mmWave system and focus on the uplink
channel estimation. A base station (BS) equipped with $N_t$ antennas serves $K$ users, each equipped with $N_r$ antennas, with the aid of an XL-RIS consisting of $M$ passive reflecting elements. All arrays are assumed to be uniform linear arrays (ULAs)\cite{IEEEhowto:zhou},\cite{IEEEhowto:songjieyang}. The carrier wavelength is denoted by $\lambda_c = c/f_c$, where $c = 3\times 10^8~\mathrm{m/s}$ is the speed of light, and the inter-element spacings at the BS, RIS, and users are set as $d_{\mathrm{BS}} = d_{\mathrm{RIS}} = d_{\mathrm{UE}} = \lambda_c/2$. Let $L$ denote the number of propagation paths of the BS-RIS channel, and let $J_k$ denote the number of propagation paths of the RIS-user channel for user $k$. The additive noise observed at the BS during the training phase is denoted by $\mathbf{N}$, whose entries are assumed to be independently and identically distributed as $\mathcal{CN}(0,\delta^2)$. For a ULA with $N$ antennas and inter-element spacing $d$, the far-field steering vector is given by
\begin{align}
\mathbf{a}(\vartheta)=\left[
1,
e^{j\frac{2\pi}{\lambda_c}d\sin\vartheta},
\ldots,
e^{j\frac{2\pi}{\lambda_c}(N-1)d\sin\vartheta}
\right]^T.
\end{align}

\subsubsection{BS-RIS Channel}
Since the BS-RIS distance is assumed to be larger than the Rayleigh distance, the BS-RIS channel is modeled under the far-field planar-wave assumption. Accordingly, the common BS-RIS channel matrix can be expressed as
\begin{align}
\mathbf{H}_{\mathrm{BR}}
&=
\sum_{l=1}^{L}
\alpha_l\,
\mathbf{a}_{N_t}\!\left(\phi_l\right)
\mathbf{a}_{M}^{H}\!\left(\omega_l\right)
\nonumber\\
&=
\mathbf{A}_{\mathrm{BS}}
\mathbf{\Lambda}
\mathbf{A}_{\mathrm{RIS}}^{H},
\end{align}
where $\alpha_l$ denotes the complex gain of the $l$th path, $\phi_l\in[0,\pi)$ and $\theta_l\in[0,\pi)$ represent the BS-side and RIS-side angles of the $l$th path, respectively, and $\mathbf{\Lambda}
=
\operatorname{diag}\!\left(
\alpha_1,\alpha_2,\ldots,\alpha_L
\right)$. Moreover, $\mathbf{A}_{\mathrm{BS}}
=\left[
\mathbf{a}_{M}\!\left(\phi_l\right),
\ldots,
\mathbf{a}_{M}\!\left(\phi_L\right)
\right]$,
$\mathbf{A}_{\mathrm{RIS}}
=
\left[
\mathbf{a}_{M}\!\left(\omega_1\right),
\ldots,
\mathbf{a}_{M}\!\left(\omega_L\right)
\right]$.

\subsubsection{RIS-User Channel}

Due to the short distance between the XL-RIS and the users, the RIS-user link generally lies in the near-field region and is therefore modeled under the spherical-wave assumption. For the $k$th user, the RIS-user channel matrix $\mathbf{H}_{\mathrm{RU},k}$ is given by
\begin{align}
\mathbf{H}_{\mathrm{RU},k}
&=
\sum_{j=1}^{J_k}
\beta_{k,j}\,
\mathbf{b}_{M}\!\left(\varphi_{k,j},r_{k,j}\right)
\mathbf{a}_{N_r}^{H}\!\left(\zeta_{k,j}\right)
\nonumber\\
&=
\mathbf{B}_{k}\,
\mathbf{\Gamma}_{k}\,
\mathbf{A}_{\mathrm{UE},k}^{H},
\end{align}
where $\beta_{k,j}$ denotes the complex gain of the $j$th path associated with user $k$, $\varphi_{k,j}\in[0,\pi)$ is the RIS-side angle, $\zeta_{k,j}\in[0,\pi)$ is the user-side angle, and $r_{k,j}$ is the distance from the RIS center to the scatterer or user location. In addition,
\begin{align}
\mathbf{\Gamma}_{k}
=
\operatorname{diag}\!\left(
\beta_{k,1},\beta_{k,2},\ldots,\beta_{k,J_k}
\right),
\end{align}
with $\mathbf{B}_{k}
=
\left[
\mathbf{b}_{M}\!\left(\varphi_{k,1},r_{k,1}\right),
\ldots,
\mathbf{b}_{M}\!\left(\varphi_{k,J_k},r_{k,J_k}\right)
\right]$,
$\mathbf{A}_{\mathrm{UE},k}
=
\left[
\mathbf{a}_{N_r}\!\left(\zeta_{k,1}\right),
\ldots,
\mathbf{a}_{N_r}\!\left(\zeta_{k,J_k}\right)
\right]$.

The near-field steering vector at the RIS is expressed as $\mathbf{b}_{M}(\varphi,r)
=
\left[
e^{-j\frac{2\pi}{\lambda_c}(r_1-r)}
\ldots,
e^{-j\frac{2\pi}{\lambda_c}(r_M-r)}
\right]^T$, where $r_m$ denotes the distance between the source/scatterer and the $m$th RIS element, given by $r_m
=
\sqrt{
r^2 +  d_m^2 - 2rd_m \sin\varphi
}$, for $ d_m
=
\left(m-\frac{M+1}{2}\right)d_{\mathrm{RIS}}$, $m=1,2,\ldots,M$.

Therefore, the considered XL-RIS-aided channel exhibits a mixed propagation structure: the BS-RIS link follows a far-field planar-wave model, whereas the RIS-user links are characterized by near-field spherical wavefronts.
%Users transmit orthogonal pilot sequences simultaneously to avoid multi-user interference. The BS receives the aggregated signal after hybrid beamforming, which is
%\begin{align}
 %\mathbf{Y} = \sum_{u=1}^K   \mathbf{W}_t^H  \mathbf{H}\mathbf{V}_t  \mathbf{h}_u   \mathbf{F}_u   \mathbf{s}_u^H  +\mathbf{W}_t^H \mathbf{N} , 
%\end{align}
%where $\mathbf{W}_t=\left [ \mathbf{U}_{N_t}\right ]_{{\left ( t-1 \right ) } N_{RF}+1:dN_{RF}}$, the $\left ( n,m \right )$-th entry of $\left [ \mathbf{U}_{N_t}\right ]$ is given by $\left [ \mathbf{U}_{N_t}\right ]_{n,m}=e^{-j\frac{2\pi \left ( n-1 \right ) \left ( m-1 \right ) }{N_{RF}} }$. N is the AWGN matrix. Leveraging the orthogonality of pilot sequences: $s_u^H s_u = \delta_{P}^2T$, $\delta_{P}$ is the transmit power, the signal dedicated to user u is separated by right-multiplying Y with  $s_u$ 
%\begin{align}
%\mathbf{Y}_{u} &= \frac{1}{\delta_{P}^2T}   \mathbf{Y} \mathbf{s}_u  \nonumber \\
%&= \mathbf{W}_{u}^{H}  \mathbf{H}\mathbf{V}_{t}  \mathbf{h}_{u}\mathbf{F}_{u} + \mathbf{N}_u,
%\end{align}
%where $\mathbf{N}_u = \frac{1}{\delta_{P}^2T} \mathbf{W}_{u}^H\mathbf{N}\mathbf{s}_u$ is the noise component for user u.

\usetikzlibrary{arrows.meta,positioning,calc,backgrounds}

\begin{figure}[t]
\centering
\resizebox{0.98\columnwidth}{!}{%
\begin{tikzpicture}[
    >=Latex,
    arr/.style={->, line width=0.65pt},
    looparr/.style={->, dashed, line width=0.65pt},
    box/.style={
        draw,
        rounded corners=1.6pt,
        line width=0.65pt,
        align=center,
        text width=2.55cm,
        minimum height=8.8mm,
        inner sep=2.1pt,
        font=\scriptsize
    },
    stageborder/.style={
        rounded corners=2.6pt,
        line width=0.95pt
    },
    stagetitle/.style={
        font=\bfseries\footnotesize,
        inner sep=1pt,
        fill=white
    }
]

% =========================================================
% Common geometry
% =========================================================
\def\StageW{11.40}
\def\StageH{5.75}
\def\Gap{1.15}

% Same three-column x-positions for both stages
\def\ColA{0.55}
\def\ColB{4.425}
\def\ColC{8.30}

% Left-column rows (4 rows)
\def\RowLone{-0.55}
\def\RowLtwo{-1.72}
\def\RowLthree{-2.89}
\def\RowLfour{-4.30}   % <-- move "typ" slightly upward

% Stage-I middle/right columns: 3 rows, vertically centered
\def\RowSone{-1.28}
\def\RowStwo{-2.45}
\def\RowSthree{-3.62}

% Stage-II rows: 3 rows
\def\RowIIone{-0.55}
\def\RowIItwo{-1.72}
\def\RowIIthree{-2.89}

% =========================================================
% Stage I coordinates
% =========================================================
\coordinate (S1NW) at (0,0);
\coordinate (S1NE) at ($(S1NW)+(\StageW,0)$);
\coordinate (S1SW) at ($(S1NW)+(0,-\StageH)$);
\coordinate (S1SE) at ($(S1NW)+(\StageW,-\StageH)$);
\coordinate (S1TopMid) at ($(S1NW)!0.5!(S1NE)$);

% =========================================================
% Stage I nodes
% =========================================================
% Left column
\node[box, fill=cyan!12, draw=cyan!60!black, anchor=north west]
(obs) at ($(S1NW)+(\ColA,\RowLone)$)
{Received signals\\
vars: $\{\mathbf{Y}_{k,t}\}$};

\node[box, fill=cyan!12, draw=cyan!60!black, anchor=north west]
(omp) at ($(S1NW)+(\ColA,\RowLtwo)$)
{User-side support estimation\\
vars: $\mathbf{Y}_{k}^{\mathrm{agg}},\ \widehat{\mathbf A}_{\mathrm{UE},k}$};

\node[box, fill=cyan!12, draw=cyan!60!black, anchor=north west]
(virt) at ($(S1NW)+(\ColA,\RowLthree)$)
{Virtual-user construction\\
vars: $\mathbf{Y}_{k,j}^{\mathrm{virt}}$};

\node[box, fill=orange!18, draw=orange!75!black, anchor=north west]
(typ) at ($(S1NW)+(\ColA,\RowLfour)$)
{Typical virtual-user selection\\
var: $\mathbf{Y}_{1}$};

% Middle column (re-centered)
\node[box, fill=blue!10, draw=blue!65!black, anchor=north west]
(ref) at ($(S1NW)+(\ColB,\RowSone)$)
{Reference-path selection\\
var: $\mathbf p_r$};

\node[box, fill=blue!10, draw=blue!65!black, anchor=north west]
(sep) at ($(S1NW)+(\ColB,\RowStwo)$)
{Path separation\\
vars: $\{\mathbf p_l\}$};

\node[box, fill=blue!10, draw=blue!65!black, anchor=north west]
(bs) at ($(S1NW)+(\ColB,\RowSthree)$)
{BS-side angle estimation\\
var: $\widehat{\mathbf A}_{\mathrm{BS}}$};

% Right column (re-centered)
\node[box, fill=blue!10, draw=blue!65!black, anchor=north west]
(polar) at ($(S1NW)+(\ColC,\RowSone)$)
{Polar sparse recovery\\
var: $\widehat{\mathbf h}_{1}^{(0)}$};

\node[box, fill=blue!10, draw=blue!65!black, anchor=north west]
(scale) at ($(S1NW)+(\ColC,\RowStwo)$)
{Other-path reconstruction\\
vars: $\widehat{\mathbf H}_{\mathrm{RIS}}$};

\node[box, fill=orange!18, draw=orange!75!black, anchor=north west]
(init) at ($(S1NW)+(\ColC,\RowSthree)$)
{Common-operator initialization\\
var: $\widehat{\mathbf S}^{(0)}$};

% Stage I arrows
\draw[arr] (obs) -- (omp);
\draw[arr] (omp) -- (virt);
\draw[arr] (virt) -- (typ);

\draw[arr] (typ.east) -- (bs.west);
\draw[arr] (bs) -- (sep);
\draw[arr] (sep) -- (ref);

\draw[arr] (ref.east) -- (polar.west);
\draw[arr] (polar) -- (scale);
\draw[arr] (scale) -- (init);

% =========================================================
% Stage II coordinates
% =========================================================
\coordinate (S2NW) at ($(S1SW)+(0,-\Gap)$);
\coordinate (S2NE) at ($(S2NW)+(\StageW,0)$);
\coordinate (S2TopMid) at ($(S2NW)!0.5!(S2NE)$);

% =========================================================
% Stage II nodes
% =========================================================
% Left column
\node[box, fill=green!12, draw=green!55!black, anchor=north west]
(allq) at ($(S2NW)+(\ColA,\RowIIone)$)
{All virtual users\\
vars: $\{\mathbf Y_q,\mathbf E_q\}$};

\node[box, fill=green!12, draw=green!55!black, anchor=north west]
(proj) at ($(S2NW)+(\ColA,\RowIItwo)$)
{BS-side projection\\
var: $\mathbf Z_q$};

\node[box, fill=green!12, draw=green!55!black, anchor=north west]
(model) at ($(S2NW)+(\ColA,\RowIIthree)$)
{Bilinear model construction\\
vars: $\mathbf z_q,\ \mathbf c_q$};

% Middle column
\node[box, fill=violet!12, draw=violet!65!black, anchor=north west]
(stepA) at ($(S2NW)+(\ColB,\RowIIone)$)
{Sparse-coefficient update\\
var: $\widehat{\mathbf c}_{q}^{(i)}$};

\node[box, fill=violet!12, draw=violet!65!black, anchor=north west]
(stepH) at ($(S2NW)+(\ColB,\RowIItwo)$)
{RIS-side channel recovery\\
var: $\widehat{\mathbf h}_{q}^{(i)}$};

\node[box, fill=violet!12, draw=violet!65!black, anchor=north west]
(stepB) at ($(S2NW)+(\ColB,\RowIIthree)$)
{Common-operator update\\
var: $\widehat{\mathbf S}^{(i+1)}$};

% ALS dashed frame and title
\coordinate (alsNW) at ($(stepA.north west)+(-0.22,0.18)$);
\coordinate (alsSE) at ($(stepB.south east)+(0.22,-0.18)$);
\coordinate (alsTR) at ($(alsSE -| alsNW)$);
\coordinate (alsTopMid) at ($(alsNW)!0.5!(alsTR)$);

\draw[violet!65!black, dashed, line width=0.95pt, rounded corners=2.6pt]
(alsNW) rectangle (alsSE);

\node[stagetitle, text=violet!60!black]
at ($(alsTopMid)+(0,0.24)$)
{ALS loop};

% Right column
\node[box, fill=green!12, draw=green!55!black, anchor=north west]
(rec) at ($(S2NW)+(\ColC,\RowIIone)$)
{BS-RIS channel recovery\\
var: $\widehat{\mathbf H}_{\mathrm{BR}}$};

\node[box, fill=green!12, draw=green!55!black, anchor=north west]
(group) at ($(S2NW)+(\ColC,\RowIItwo)$)
{Physical-user regrouping\\
var: $\widehat{\mathbf H}_{\mathrm{RU},k}$};

\node[box, fill=orange!18, draw=orange!75!black, anchor=north west]
(final) at ($(S2NW)+(\ColC,\RowIIthree)$)
{Cascaded-channel reconstruction\\
var: $\widehat{\mathbf G}_{k,n}$};

% Stage II arrows
\draw[arr] (allq) -- (proj);
\draw[arr] (proj) -- (model);

\draw[arr] (model.east) -- (stepA.west);
\draw[arr] (stepA) -- (stepH);
\draw[arr] (stepH) -- (stepB);
\draw[looparr] (stepB.east) .. controls +(6mm,0) and +(6mm,0) .. (stepA.east);

\draw[arr] (stepB.east) -- (rec.west);
\draw[arr] (rec) -- (group);
\draw[arr] (group) -- (final);

% Connections between stages
\draw[arr] (typ.south) -- ++(0,-5mm) -| (allq.north);
\draw[arr] (init.south) -- ++(0,-5mm) -| (stepA.north);

% =========================================================
% Stage II bottom tightly fitted to content
% =========================================================
\coordinate (S2BottomRef) at ($(alsSE)+(0,-0.40)$);
\coordinate (S2SW) at ($(S2NW |- S2BottomRef)$);
\coordinate (S2SE) at ($(S2NE |- S2BottomRef)$);

% =========================================================
% Draw stage frames on background layer
% =========================================================
\begin{scope}[on background layer]
    \draw[stageborder, draw=cyan!60!black] (S1NW) rectangle (S1SE);
    \draw[stageborder, draw=green!55!black] (S2NW) rectangle (S2SE);
\end{scope}

% Stage titles
\node[stagetitle, text=cyan!45!black]
at ($(S1TopMid)+(0,0.34)$)
{Stage I: Typical-user initialization};

\node[stagetitle, text=green!40!black]
at ($(S2TopMid)+(0,0.34)$)
{Stage II: Joint refinement};

\end{tikzpicture}%
}
\caption{The channel estimation process of the proposed method.}
\label{fig:xlris_flowchart_centered_shifted}
\end{figure}
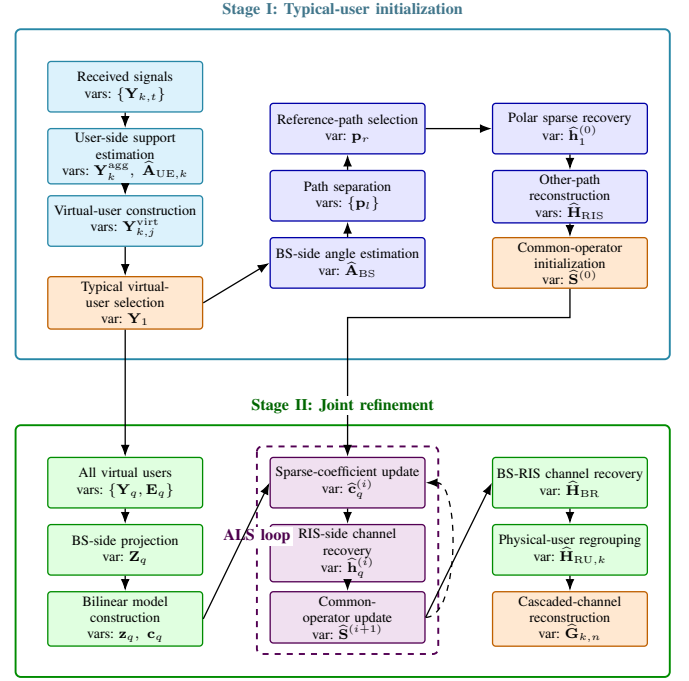

\section{Channel Estimation}

\subsection{Channel Estimation for the Typical User}

\subsubsection{User-Side Angular Support Estimation and Virtual-User Construction}

For the $k$th user, let $\mathbf{Y}_{k,t}\in\mathbb{C}^{N_t\times N_r}$ denote the received signal matrix at the BS during the $t$th RIS training slot, i.e.,
\begin{align}
\mathbf{Y}_{k,t}
=
\sqrt{p}\,\mathbf{H}_{\mathrm{BR}}
\operatorname{Diag}(\mathbf{e}_t)\mathbf{H}_{\mathrm{RU},k}
+\mathbf{N}_{k,t},
\end{align}
where $t=1,2,\ldots,\tau$, $\mathbf{e}_t\in\mathbb{C}^{M}$ is the RIS phase vector at slot $t$.
By stacking the transposed received matrices over all training slots, we form the aggregated observation $\mathbf{Y}_{k}^{\mathrm{agg}}
=
\big[
\mathbf{Y}_{k,1}^{T},
\mathbf{Y}_{k,2}^{T},
\ldots,
\mathbf{Y}_{k,\tau}^{T}
\big]
\in\mathbb{C}^{N_r\times N_t\tau}$. Using the channel model $\mathbf{H}_{\mathrm{RU},k}
=
\mathbf{B}_k\mathbf{\Gamma}_k\mathbf{A}_{\mathrm{UE},k}^{H}$, the aggregated observation can be written as $\mathbf{Y}_{k}^{\mathrm{agg}}
=
\mathbf{A}_{\mathrm{UE},k}\mathbf{\Xi}_k
+
\mathbf{N}_{k}^{\mathrm{agg}}$,
where $\mathbf{\Xi}_k\in\mathbb{C}^{J_k\times N_t\tau}$ collects the effective path-dependent coefficients across all training slots, and thus it becomes row-sparse with respect to the user-side angular domain. To estimate the user-side support, we introduce an overcomplete user-side dictionary $\mathbf{A}_{\mathrm{U}}
=
\big[
\mathbf{a}_{N_r}(\vartheta_1),
\mathbf{a}_{N_r}(\vartheta_2),
\ldots,
\mathbf{a}_{N_r}(\vartheta_D)
\big]$. Then we obtain 
\begin{align}
\mathbf{Y}_{k}^{\mathrm{agg}}
=
\mathbf{A}_{\mathrm{U}}\widetilde{\mathbf{\Xi}}_k
+
\mathbf{N}_{k}^{\mathrm{agg}},
\label{eq:Yagg_sparse_dic}
\end{align}
where $\widetilde{\mathbf{\Xi}}_k\in\mathbb{C}^{D\times N_t\tau}$ is a row-sparse matrix with $J_k$ dominant nonzero rows. Therefore, the user-side support can be recovered via OMP, yielding the support set $\widehat{\Omega}_k$ and the corresponding estimated user-side steering matrix $\widehat{\mathbf{A}}_{\mathrm{UE},k}
=
\mathbf{A}_{\mathrm{U}}(:,\widehat{\Omega}_k)$.

With $\widehat{\mathbf{A}}_{\mathrm{UE},k}$, the multi-antenna channel of user $k$ is decomposed into $\widehat{J}_k=|\widehat{\Omega}_k|$ virtual single-antenna components. Specifically, for each training slot $t$, we perform user-side subspace separation as
\begin{align}
\widetilde{\mathbf{Y}}_{k,t}
=
\mathbf{Y}_{k,t}\widehat{\mathbf{A}}_{\mathrm{UE},k}
\left(
\widehat{\mathbf{A}}_{\mathrm{UE},k}^{H}
\widehat{\mathbf{A}}_{\mathrm{UE},k}
\right)^{-1}
\in
\mathbb{C}^{N_t\times \widehat{J}_k}.
\label{eq:virtual_sep}
\end{align}

The $j$th column of $\widetilde{\mathbf{Y}}_{k,t}$ corresponds to the $j$th virtual user of physical user $k$. Stacking these columns over all RIS training slots yields the observation matrix of the virtual user $(k,j)$ as
\begin{align}
\mathbf{Y}_{k,j}^{\mathrm{virt}}
=
\big[
\widetilde{\mathbf{Y}}_{k,1}(:,j),
\widetilde{\mathbf{Y}}_{k,2}(:,j),
\ldots,
\widetilde{\mathbf{Y}}_{k,\tau}(:,j)
\big]
\in\mathbb{C}^{N_t\times \tau}.
\label{eq:virtual_user_obs}
\end{align}

Without loss of generality, we select one virtual component extracted from user 1 as the typical virtual user and denote its observation by $\mathbf{Y}_1=\mathbf{Y}_{1,1}^{\mathrm{virt}}$.
The subsequent estimation of the common BS-side angular parameters and the RIS-side cascaded channel is then carried out based on $\mathbf{Y}_1$.

\subsubsection{BS-Side Parameter Estimation From the Typical Virtual User}

We select one virtual component extracted from user~1 as the typical virtual user. Let its observation matrix be denoted by $\mathbf{Y}_1 \in \mathbb{C}^{N_t\times \tau}$.

The BS-side angles are common to all virtual users through the shared BS-RIS channel. Hence, they can be estimated from $\mathbf{Y}_1$ via a DFT-based procedure with angle-rotation compensation in \cite{IEEEhowto:peng}. Let $\widehat{L}$ denote the number of detected BS-RIS paths, the corresponding BS-side angle is obtained as $\widehat{\phi}_l
=
\arcsin\!\left(
\frac{\lambda_c}{d_{\mathrm{BS}}}\widehat{\psi}_l
\right)$. Accordingly, the estimated BS steering matrix is constructed as $\widehat{\mathbf{A}}_{\mathrm{BS}}
=
\big[
\mathbf{a}_{N_t}(\widehat{\phi}_1),
\mathbf{a}_{N_t}(\widehat{\phi}_2),
\ldots,
\mathbf{a}_{N_t}(\widehat{\phi}_{\widehat{L}})
\big]$.

Next, $\mathbf{Y}_1$ is projected onto the estimated BS-side angular subspace to separate the common BS-RIS paths
\begin{align}
\mathbf{p}_l
&=
\frac{1}{N_t\sqrt{p}}
\widehat{\mathbf{A}}_{\mathrm{BS}}^{H}\mathbf{Y}_1
\nonumber\\
&=
\big[
\mathbf{p}_1,\mathbf{p}_2,\ldots,\mathbf{p}_{\widehat{L}}
\big]^{H}
+
\widetilde{\mathbf{N}}_1,
\label{eq:AoA_projection_virtual}
\end{align}
where $\mathbf{p}_l\in\mathbb{C}^{\tau}$ denotes the observation vector associated with the $l$th separated BS--RIS path.

Among the separated paths, we select the dominant one as the reference path by using $r
=
\arg\max_{1\le l\le \widehat{L}}
\|\mathbf{p}_l\|_2^2$.
For notational compactness, let $\omega_l$ denote the RIS-side spatial frequency corresponding to the $l$th BS--RIS path. According to \cite{IEEEhowto:songjieyang}, since the positions of BS and RIS are fixed, the angle information of the LoS path between BS and RIS can be directly obtained. Then the reference path is associated with the known $\omega_r$.  To recover the RIS-side channel of the typical virtual user, we employ the near-field polar-domain dictionary $\mathbf{P}_{\mathrm{bm}}\in\mathbb{C}^{M\times D}$ introduced in \cite{IEEEhowto:Dai}. Using the reference path $\omega_r$, the corresponding compensated sensing matrix is constructed as $\mathbf{D}_r
=
\mathbf{E}_1^{H}
\operatorname{Diag}\!\big\{
\mathbf{a}_{M}(\omega_r)
\big\}
\mathbf{P}_{\mathrm{bm}}$, where $\mathbf{E}_1$ denotes the RIS training matrix associated with the typical virtual user. Consequently, the observation corresponding to the reference path admits the sparse linear model
\begin{align}
\mathbf{p}_r
=
\mathbf{D}_r\boldsymbol{\gamma}_1
+\mathbf{n}_r,
\label{eq:pr_sparse_virtual}
\end{align}
where $\boldsymbol{\gamma}_1\in\mathbb{C}^{D}$ is a sparse coefficient vector. It can be estimated via LAOMP as
\begin{align}
\widehat{\boldsymbol{\gamma}}_1
=
\arg\min_{\boldsymbol{\gamma}_1}
\left\|
\mathbf{p}_r-\mathbf{D}_r\boldsymbol{\gamma}_1
\right\|_2^2
+
\lambda\|\boldsymbol{\gamma}_1\|_0.
\label{eq:laomp_virtual}
\end{align}

Let $\Omega_1$ denote the support set of $\widehat{\boldsymbol{\gamma}}_1$. Then the equivalent RIS-side channel vector of the typical virtual user is reconstructed as $\widehat{\mathbf{h}}_1^{(0)}
=
\mathbf{P}_{\mathrm{bm}}\widehat{\boldsymbol{\gamma}}_1$, which provides an initialization for the subsequent joint refinement. Based on $\widehat{\mathbf{h}}_1^{(0)}$, the cascaded vector corresponding to the reference path is given by
\begin{align}
\widehat{\mathbf{h}}_{\mathrm{RIS},r}
=
\operatorname{Diag}\!\big\{
\mathbf{a}_{M}(\omega_r)
\big\}
\widehat{\mathbf{h}}_1^{(0)}.
\label{eq:hRISr_virtual}
\end{align}

To avoid constructing an additional dictionary for every remaining BS--RIS path, we exploit the angle-gain scaling property. Specifically, for $l\neq r$, the RIS-side cascaded vector corresponding to the $l$th path can be represented as
\begin{align}
\mathbf{h}_{\mathrm{RIS},l}
=
\operatorname{Diag}\!\big\{
\mathbf{a}_{M}(\Delta\omega_l)
\big\}
\mathbf{h}_{\mathrm{RIS},r}\eta_l,
\label{eq:angle_gain_scaling_virtual}
\end{align}
where $\Delta\omega_l=\omega_l-\omega_r$, and $\eta_l\in\mathbb{C}$ is the complex scaling coefficient between the $l$th path and the reference path. Based on $\widehat{\mathbf{h}}_{\mathrm{RIS},r}$, define
\begin{align}
\mathbf{z}_l(\Delta\omega)
=
\mathbf{E}_1^{H}
\operatorname{Diag}\!\big\{
\widehat{\mathbf{h}}_{\mathrm{RIS},r}
\big\}
\mathbf{a}_{M}(\Delta\omega).
\label{eq:z_l_virtual}
\end{align}

Then $\Delta\omega_l$ is estimated via correlation maximization as
\begin{align}
\Delta\widehat{\omega}_l
=
\arg\max_{\Delta\omega\in
\left[
-2\frac{d_{\mathrm{RIS}}}{\lambda_c},
\,
2\frac{d_{\mathrm{RIS}}}{\lambda_c}
\right]}
\left|
\left\langle
\mathbf{p}_l,\mathbf{z}_l(\Delta\omega)
\right\rangle
\right|,
\qquad l\neq r.
\label{eq:Deltaomega_hat_virtual}
\end{align}

Given $\Delta\widehat{\omega}_l$, the corresponding scaling coefficient is estimated by least squares as $\widehat{\eta}_l
=
\left(
\mathbf{z}_l^{H}(\Delta\widehat{\omega}_l)
\mathbf{z}_l(\Delta\widehat{\omega}_l)
\right)^{-1}
\mathbf{z}_l^{H}(\Delta\widehat{\omega}_l)\mathbf{p}_l$. Hence, the RIS-side cascaded vector associated with the $l$th path is reconstructed as $\widehat{\mathbf{h}}_{\mathrm{RIS},l}
=
\operatorname{Diag}\!\big\{
\mathbf{a}_{M}(\Delta\widehat{\omega}_l)
\big\}
\widehat{\mathbf{h}}_{\mathrm{RIS},r}\widehat{\eta}_l,
l\neq r$.

Collecting all separated BS-RIS paths, the initial cascaded channel corresponding to the typical virtual user can be expressed as
\begin{align}
\widehat{\mathbf{G}}_1^{(0)}
=
\widehat{\mathbf{A}}_{\mathrm{BS}}
\widehat{\mathbf{H}}_{\mathrm{RIS}}^{H},
\label{eq:G1_hat_virtual}
\end{align}
where $\widehat{\mathbf{H}}_{\mathrm{RIS}}
=
\big[
\widehat{\mathbf{h}}_{\mathrm{RIS},1},
\widehat{\mathbf{h}}_{\mathrm{RIS},2},
\ldots,
\widehat{\mathbf{h}}_{\mathrm{RIS},\widehat{L}}
\big]$. Moreover, using $\widehat{\mathbf{h}}_1^{(0)}$, we can initialize the common BS--RIS operator for the subsequent ALS refinement. Define
\begin{align}
\mathbf{Z}_1
&=
\frac{1}{N_t\sqrt{p}}
\widehat{\mathbf{A}}_{\mathrm{BS}}^{H}\mathbf{Y}_1,
\nonumber \\
\mathbf{X}_1^{(0)}
&=
\operatorname{Diag}\!\big\{
\widehat{\mathbf{h}}_1^{(0)}
\big\}
\mathbf{E}_1.
\end{align}

Then the initial estimate is obtained as
\begin{align}
\mathbf{S}^{(0)}
=
\mathbf{Z}_1
\big(\mathbf{X}_1^{(0)}\big)^{H}
\left(
\mathbf{X}_1^{(0)}
\big(\mathbf{X}_1^{(0)}\big)^{H}
\right)^{\dagger}.
\label{eq:S_init_virtual}
\end{align}

\subsection{Channel Estimation for Other Virtual Users and Joint Refinement for All Virtual Users}

Let $K_{\mathrm{v}}=\sum_{k=1}^{K}\widehat{J}_k$ denote the total number of virtual users obtained after the user-side decomposition stage. For notational simplicity, we re-index all virtual users by a single index $q\in\{1,2,\ldots,K_{\mathrm{v}}\}$. For the $q$th virtual user, the effective training observation can be written as
\begin{equation}
\mathbf{Y}_q
=
\sqrt{p}\,\mathbf{H}_{\mathrm{BR}}
\operatorname{Diag}(\mathbf{h}_q)\mathbf{E}_q
+\mathbf{N}_q,
\label{eq:Yq_effective_virtual}
\end{equation}
where $\mathbf{h}_q\in\mathbb{C}^{M}$ denotes the equivalent RIS-side channel vector of the $q$th virtual user, and $\mathbf{E}_q$ is the corresponding RIS training matrix.

Recalling the far-field decomposition of the BS-RIS channel $\mathbf{H}_{\mathrm{BR}}=\mathbf{A}_{\mathrm{BS}}\mathbf{\Lambda}\mathbf{A}_{\mathrm{RIS}}^{H}$ and using the approximation $\widehat{\mathbf{A}}_{\mathrm{BS}}^{H}\mathbf{A}_{\mathrm{BS}}
\approx
N_t\mathbf{I}_{L}$, we project $\mathbf{Y}_q$ onto the estimated BS-side angular subspace by using $\mathbf{Z}_q
=
\frac{1}{N_t\sqrt{p}}
\widehat{\mathbf{A}}_{\mathrm{BS}}^{H}\mathbf{Y}_q$ and obtain
\begin{align}
\mathbf{Z}_q&\approx
\underbrace{
\left(
\frac{1}{N_t}
\widehat{\mathbf{A}}_{\mathrm{BS}}^{H}\mathbf{H}_{\mathrm{BR}}
\right)
}_{\mathbf{S}}
\operatorname{Diag}(\mathbf{h}_q)\mathbf{E}_q
+\widetilde{\mathbf{N}}_q,
\label{eq:Zq_projection_virtual}
\end{align}
where $\widetilde{\mathbf{N}}_q=\frac{1}{N_t\sqrt{p}}\widehat{\mathbf{A}}_{\mathrm{BS}}^{H}\mathbf{N}_q$, and $\mathbf{S}$ is given by 
\begin{align}
\mathbf{S}
=
\mathbf{\Lambda}\mathbf{A}_{\mathrm{RIS}}^{H}.
\end{align}

The operator $\mathbf{S}$ absorbs all common BS--RIS parameters after BS-side projection. Consequently, the non-separable coupling introduced by the cascaded channel is significantly simplified. Moreover, the model in \eqref{eq:Zq_projection_virtual} is bilinear and thus has an inherent scale ambiguity: multiplying $\mathbf{S}$ by a nonzero scalar and dividing $\mathbf{h}_q$ by the same scalar does not change the final cascaded channel estimate. By vectorizing $\mathbf{z}_q$, we obtain
\begin{align}
\mathbf{z}_q
=
\left(
\mathbf{E}_q^{T}\diamond\mathbf{S}
\right)\mathbf{h}_q
+\mathbf{n}_q,
\label{eq:vec_model_virtual}
\end{align}
where $\diamond$ denotes the Khatri--Rao product.

To exploit the near-field sparsity of $\mathbf{h}_q$, we further represent it over the near-field polar-domain dictionary. Specifically $\mathbf{h}_q
=
\mathbf{P}_{\mathrm{bm}}\mathbf{c}_q$,
where $\mathbf{c}_q$ is a sparse coefficient vector satisfying $\|\mathbf{c}_q\|_0\le J_q$. Substituting (21) into (22) yields
\begin{equation}
\mathbf{z}_q
=
\underbrace{
\left(
\mathbf{E}_q^{T}\diamond\mathbf{S}
\right)\mathbf{P}_{\mathrm{bm}}
}_{\boldsymbol{\Psi}_q(\mathbf{S})}
\mathbf{c}_q
+\mathbf{n}_q.
\label{eq:zq_sparse_model_virtual}
\end{equation}

To improve robustness, we jointly refine $\mathbf{S}$ using the observations of all virtual users. This leads to the following bilinear sparse optimization problem:
\begin{align}
\min_{\mathbf{S},\,\{\mathbf{c}_q\}}
\quad
&\sum_{q=1}^{K_{\mathrm{v}}}
\left\|
\mathbf{Z}_q
-
\mathbf{S}\operatorname{Diag}(\mathbf{P}_{\mathrm{bm}}\mathbf{c}_q)\mathbf{E}_q
\right\|_F^{2}
\nonumber\\
\text{s.t.}\quad
&\|\mathbf{c}_q\|_0 \le J_q,\qquad q=1,2,\ldots,K_{\mathrm{v}}.
\label{eq:joint_opt_virtual}
\end{align}

Since \eqref{eq:joint_opt_virtual} is bilinear in $\mathbf{S}$ and $\{\mathbf{c}_q\}$, we adopt an alternating least-squares (ALS) procedure for iterative refinement.

\subsubsection{Step A: Estimation of Virtual-User Sparse Coefficients}

For a fixed $\mathbf{S}^{(i)}$, the sparse coefficients of each virtual user can be estimated independently by solving
\begin{equation}
\widehat{\mathbf{c}}_q^{(i)}
=
\arg\min_{\mathbf{c}_q}
\left\|
\mathbf{z}_q
-
\boldsymbol{\Psi}_q\!\left(\mathbf{S}^{(i)}\right)\mathbf{c}_q
\right\|_2^{2}
\quad
\text{s.t.}\ 
\|\mathbf{c}_q\|_0\le J_q,
\label{eq:ALS_stepA_virtual}
\end{equation}
which can be efficiently handled by LAOMP. The corresponding channel estimate is reconstructed as $\widehat{\mathbf{h}}_q^{(i)}
=
\mathbf{P}_{\mathrm{bm}}\widehat{\mathbf{c}}_q^{(i)}$.

\subsubsection{Step B: Update of the Common Operator}

For fixed $\{\widehat{\mathbf{h}}_q^{(i)}\}_{q=1}^{K_{\mathrm{v}}}$, define $\mathbf{X}_q^{(i)}
=
\operatorname{Diag}\!\left(\widehat{\mathbf{h}}_q^{(i)}\right)\mathbf{E}_q$, then the update of $\mathbf{S}$ is obtained by solving a least-squares problem, whose closed-form solution is
\begin{equation}
\mathbf{S}^{(i+1)}
=
\left(
\sum_{q=1}^{K_{\mathrm{v}}}
\mathbf{Z}_q\left(\mathbf{X}_q^{(i)}\right)^{H}
\right)
\left(
\sum_{q=1}^{K_{\mathrm{v}}}
\mathbf{X}_q^{(i)}\left(\mathbf{X}_q^{(i)}\right)^{H}
\right)^{\dagger}.
\label{eq:ALS_stepB_virtual}
\end{equation}

The above two steps are repeated until convergence or until a prescribed maximum number of iterations is reached. After $I$ ALS iterations, the BS--RIS channel is reconstructed as $\widehat{\mathbf{H}}_{\mathrm{BR}}
=
\widehat{\mathbf{A}}_{\mathrm{BS}}\mathbf{S}^{(I)}$.
Next, the virtual-user estimates are regrouped to recover the physical-user channels. Let $\mathcal{V}_k$ denote the index set of virtual users associated with physical user $k$, and let $|\mathcal{V}_k|=\widehat{J}_k$. Construct
\begin{align}
\widehat{\mathbf{H}}_{\mathrm{v},k}
=
\big[
\widehat{\mathbf{h}}_{q_1}^{(I)},
\widehat{\mathbf{h}}_{q_2}^{(I)},
\ldots,
\widehat{\mathbf{h}}_{q_{\widehat{J}_k}}^{(I)}
\big],
\qquad q_j\in\mathcal{V}_k,
\label{eq:Hv_k_virtual}
\end{align}
from which the RIS--user channel of physical user $k$ is reconstructed as $\widehat{\mathbf{H}}_{\mathrm{RU},k}
=
\widehat{\mathbf{H}}_{\mathrm{v},k}\widehat{\mathbf{A}}_{\mathrm{UE},k}^{H}$. Finally, for the $n$th receive antenna of user $k$, the corresponding cascaded channel is obtained as
\begin{align}
\widehat{\mathbf{G}}_{k,n}
=
\widehat{\mathbf{H}}_{\mathrm{BR}}
\operatorname{Diag}\!\left(
[\widehat{\mathbf{H}}_{\mathrm{RU},k}]_{:,n}
\right),
\qquad n=1,2,\ldots,N_r.
\label{eq:Gkn_final_virtual}
\end{align}

Equivalently, the cascaded channel of user $k$ can be represented by the collection
\begin{align}
\widehat{\mathbf{G}}_k
=
\left\{
\widehat{\mathbf{G}}_{k,n}
\right\}_{n=1}^{N_r}. 
\end{align}

The entire estimation process is illustrated in Fig. 1, including the estimation for the typical user and the subsequent joint optimization. The figure also provides the specific variable names for the relevant formulas involved.

\subsection{Pilot Overhead and Complexity Analysis}

Let the number of paths between all users and RIS be the same, $D_U$ and $D_P$ denote the user-side angular dictionary size and the RIS-side polar dictionary size, respectively. Since $D_U \propto N_r$ and $D_P \propto M$, the required pilot overhead is
$T_{\mathrm{prop}} \approx \mathcal{O}\!\left(KJ\log N_r + J\log M + \frac{(K-1)J\log M}{L}\right)-KJ$,
 By contrast, the benchmark in \cite{IEEEhowto:songjieyang} scales as $T_{\mathrm{ref}}=K(1+QLJ)$ when $T=K$. $Q$ is the number of phase adjustments.

Let $K_v=\sum_{k=1}^{K}\hat J_k$ denote the total number of virtual users and let $I_{\rm ALS}$ denote the number of ALS iterations. In each ALS iteration, Step A updates the sparse coefficients of all $K_v$ virtual users via LAOMP, while Step B updates the common operator $\mathbf{S}$ by a closed-form LS solution. Therefore, the per-iteration complexity of the ALS refinement can be expressed as $\mathcal{O}\!\left(K_v\mathcal{C}_{\rm A}+\mathcal{C}_{\rm B}\right)$,
where $\mathcal{C}_{\rm A}$ and $\mathcal{C}_{\rm B}$ denote the complexities of one sparse recovery in Step A and one LS update in Step B, respectively. Hence, the total complexity is $\mathcal{O}\!\left(\mathcal{C}_{\rm init}+I_{\rm ALS}\left(K_v\mathcal{C}_{\rm A}+\mathcal{C}_{\rm B}\right)\right)$. Similar to \cite{IEEEhowto:songjieyang}, the dominant computational burden mainly comes from sparse recovery, especially when the number of required iterations is very small. This can be proved in Fig. 2. Therefore, the increase in complexity due to a small number of iterations is very limited.

\section{Simulation Results}
The carrier frequency of the system is 30 GHz with a
bandwidth of 100 MHz. The BS is equipped with $N=128$ antennas, the RIS has $M=256$ reflecting elements, and each user is equipped with $N_r=32$ antennas. The number of users is $K=4$, the number of BS-RIS paths is fixed as $L=3$. The complex channel gain is generated as $\alpha_l \sim \mathcal{CN}\!\left(0,\,10^{-3} d_{\mathrm{BR}}^{-2.8}\right)$ and $\beta_{k,j} \sim \mathcal{CN}\!\left(0,\,10^{-3} d_{\mathrm{RU}}^{-2.2}\right)$, where $d_{\mathrm{BR}} = 100~\mathrm{m}$ and $d_{\mathrm{RU}} \in \left [ 1,50 \right ] ~\mathrm{m}$. $\mathrm{SNR} = 10\log\!\left(10^{-6} d_{\mathrm{BR}}^{-2.2} d_{\mathrm{RU}}^{-2.8} p / \delta^2\right)$. The
transmit power for all users is set to $p = 1~\mathrm{W}$. To evaluate the
performance of each method, the normalized mean square error($\mathrm{NMSE}$) is adopted and defined as $\mathrm{NMSE}=\mathbb{E} ( \left\| {\textstyle \sum_{k=1}^{K}} \hat{\mathbf{G}}_{k}-\mathbf{G}_{k}\right\|_F^2 ) /
\mathbb{E}( \sum_{k=1}^{K}\left\|\mathbf {G}_{k}\right\|_F^2 ) $.

\begin{figure}[!t]
    \centering
    \includegraphics[width=1.0\columnwidth]{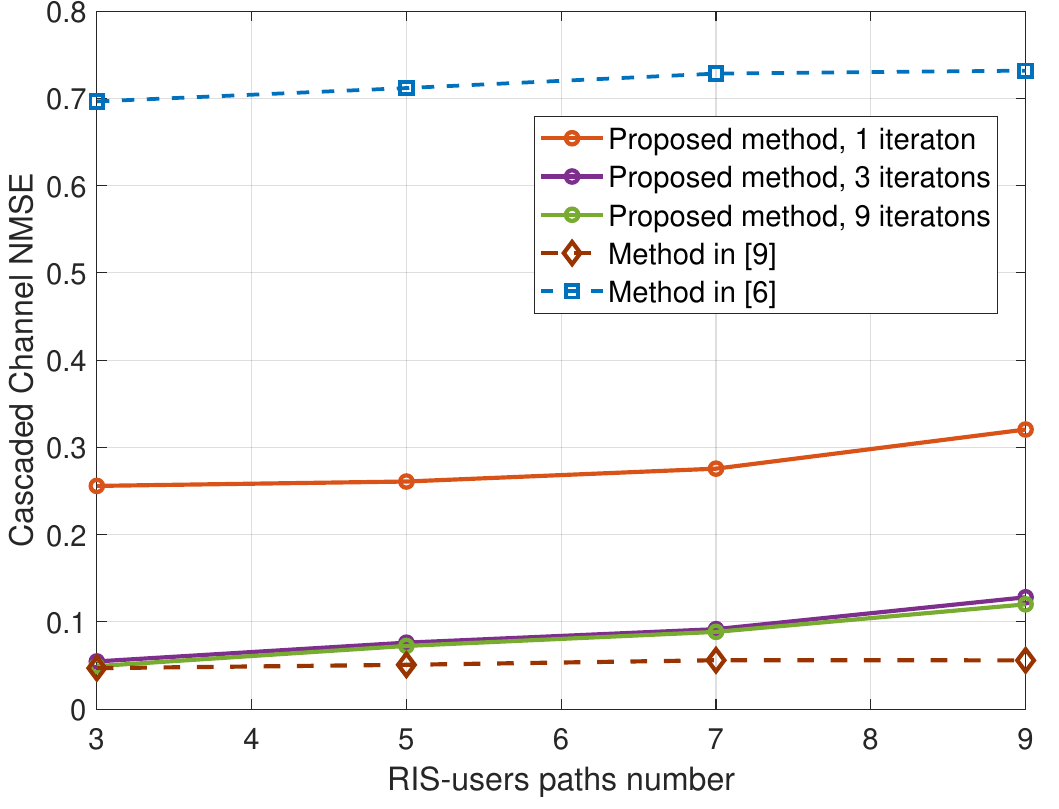}
    \caption{NMSEs vs paths number between RIS and the users.}
    \label{fig:fig1}
\end{figure}

\begin{figure}[!t]
    \centering
    \includegraphics[width=1.0\columnwidth]{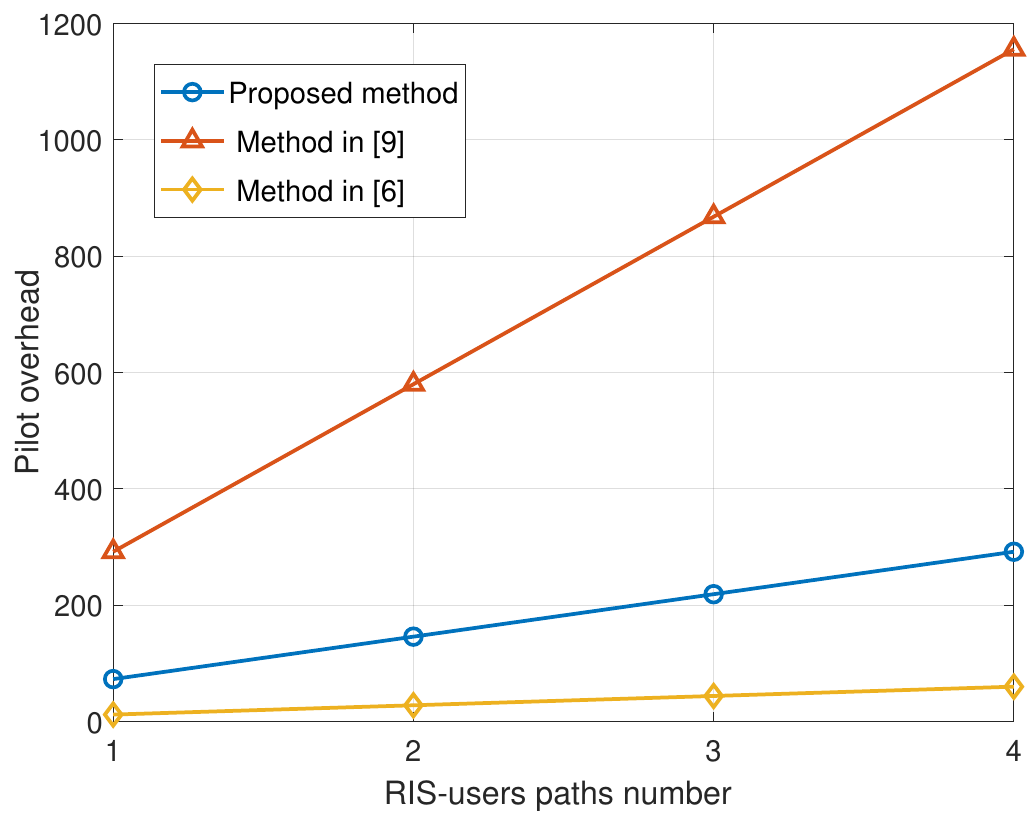}
    \caption{Pilot overhead vs paths number between RIS and the users.}
    \label{fig:fig1}
\end{figure}
Fig. 2 shows the cascaded channel NMSE versus the number of RIS-user paths at SNR =10 dB. The proposed scheme with ALS refinement consistently achieves much lower NMSE than the one-shot initialization. Increasing the ALS iterations from 1 to 3 brings a clear performance gain, while the improvement from 3 to 9 iterations is only marginal, indicating that the proposed bilinear refinement converges quickly. Although channel estimation becomes more difficult as 
paths number increases, the refined proposed scheme remains competitive with the benchmark in \cite{IEEEhowto:songjieyang}, which verifies its effectiveness for near-field cascaded channel estimation. We also evaluate the far-field estimation scheme under the near-field spherical-wave model \cite{IEEEhowto:zhou}. The results show that the far-field scheme suffers from severe degradation in this case, further highlighting the necessity of near-field channel estimation.

Fig. 3 compares the pilot overhead of the proposed scheme with that of the far-field method and the benchmark methods, where $Q=24$. It is evident that the proposed method requires substantially fewer pilots. This advantage stems from the fact that the proposed pilot overhead scales only logarithmically with the  dictionary size, whereas the benchmark method still exhibits a linear dependence on the number of paths.  As a result, the gap in pilot overhead enlarges progressively as the channel becomes more complex. These results verify that the proposed framework can effectively exploit the common BS-RIS structure and the near-field sparsity and multi-antenna users, thereby achieving a more scalable channel acquisition strategy for XL-RIS-aided systems.

\section{Conclusion}
This paper studied cascaded channel estimation for near-field XL-RIS-aided MIMO systems. By leveraging the common BS-RIS channel characteristics and the polar-domain sparsity of the RIS-user links, a low-overhead two-stage estimation framework was proposed. The method firstly constructs virtual single-antenna users to extract the common BS-side parameters and initialize the RIS-side channels, and then employs an ALS-based refinement to jointly update the common operator and user-specific channel components. Simulation results showed that the proposed scheme can achieve competitive NMSE performance with substantially lower pilot overhead than existing benchmarks, while maintaining good robustness as the number of RIS-user paths increases. Therefore, the proposed design solution provides an efficient and scalable solution for channel estimation in actual near-field XL-RIS-assisted millimeter-wave communication.
%\section*{Acknowledgment}


\ifCLASSOPTIONcaptionsoff
  \newpage
\fi



%\bibliographystyle{IEEEtran}
\bibliographystyle{IEEEtran}
\bibliography{ref}


%\vfill


\end{document}